\documentclass[11pt,letterpapert]{article}
\usepackage{makeidx}         
\usepackage{graphicx}        
\usepackage{multicol}        
\usepackage[bottom]{footmisc}

\usepackage{amssymb}
\usepackage{epstopdf}
\usepackage{fancyhdr}
\usepackage{amsmath}
\usepackage{hyperref}
\usepackage{color}
\usepackage{soul}

\usepackage{enumerate}
\definecolor{purple}{rgb}{.9,0,.9}

\usepackage{geometry}
\def\Nat{{\, \hbox{N \kern-1.25em I}\ \;}}
\def\Real{{\, \hbox{R \kern-1.25em I}\ \;}}
\def\Pos{{\, \hbox{P \kern-1.15em I}\ \;}}

\def\Int{{\, \hbox{\tenss Z \kern-1.1em Z} \,}}

\def\rfa{\qquad {\rm for \ all}\ \ }

 \def\cS{{\cal S}}
\def\cW{{\cal W}}

\def\bd{{\bf d}}
\def\be{{\bf e}}\def\bff{{\bf f}}

\def\bn{{\bf n}}
\def\br{{\bf r}}

\def\bC{{\bf C}}

\def\bL{{\bf L}}

\def\b0{{\bf 0}}

\def\ddd{d}

\def\beqn{\begin{equation}}
\def\eeqn{\end{equation}}

\def\Nat{\mathbb{N}}

\def\Real{\mathbb{R}}
\def\Pos{\mathbb{P}}

\def\Int{{\, \hbox{\tenss Z \kern-1.1em Z} \,}}

\def\hr{\hat\br}

\newcommand{\da}{\,\text{d}a_y}

\newcommand{\dr}{\,\text{d}r}

\newcommand{\half}{\frac{1}{2}}

\newcommand{\abs}[1]{|#1\rvert}

\title{Calculating the bending moduli of the Canham--Helfrich free-energy density from a particular potential}
\author{Brian Seguin$^*$ \& Eliot Fried$^{**}$
\\[4pt]
{\small $^*$Division of Mathematics}
\\[-2pt]
{\small University of Dundee}
\\[-2pt]
{\small Dundee DD1 4HN}
\\[-2pt]
{\small Scotland, UK}
\\[4pt]
{\small $^{**}$Mathematical Soft Matter Unit}
\\[-2pt]
{\small Okinawa Institute of Science and Technology}
\\[-2pt]
{\small 1919-1 Tancha, Onna-son, Kunigami-gun}
\\[-2pt]
{\small Okinawa, Japan 904-0495}}

\begin{document}
\date{}

\maketitle

\begin{abstract}
\noindent The Canham--Helfrich free-energy density for a lipid bilayer involves the mean and Gaussian curvatures of the midsurface of the bilayer. The splay and saddle-splay moduli $\kappa$ and $\bar\kappa$ regulate the sensitivity of the free-energy density to changes of these curvatures. Seguin and  Fried derived the Canham--Helfrich energy by taking into account the interactions between the molecules comprising the bilayer, giving rise to integral representations for the moduli in terms of the interaction potential. In the present work, two potentials are chosen and the integrals are evaluated to yield expressions for the moduli, which are found to depend on parameters associated with each potential.  These results are compared with values of the moduli found in the current literature.
\\[8pt] \textbf{Keywords}: interaction potential; lipid molecules; splay modulus; saddle-splay modulus; biomembrane; vesicle; curvature elasticity
\end{abstract}

\section{Introduction}

Biomembranes are ubiquitous in nature. An essential element of a biomembrane is a lipid bilayer, which is composed of phospholipid molecules. These molecules have hydrophilic head groups and a pair of hydrophobic tails. Due to these properties, when a large number of lipid molecules are placed in a solution, they self-assemble, under suitable conditions, into two-dimensional structures consisting of two leaflets (or monolayers). The lipid molecules are oriented so that the tails of the molecules in each leaflet are in contact with each other, while the head groups are in contact with the suspending solution; see, for example, Lasic~\cite{L}. These two-dimensional structures often close to form vesicles and are usually between 50 nanometers and tens of micrometers in diameter but only a few nanometers thick, as observed by Luisi and Walde~\cite{LW}. Due to these dimensions, lipid bilayers are usually modeled as surfaces.

An accepted expression for the free-energy density of a lipid bilayer takes the form
\beqn
\psi=2\kappa(H-H_\circ)^2+\bar\kappa K,
\label{CHexpression}
\eeqn
where $H$ and $K$ denote the mean and Gaussian curvatures of the midsurface of the bilayer, $\kappa$ and $\bar\kappa$ are the splay and saddle-splay moduli, respectively, and $H_\circ$ is the spontaneous mean-curvature, which describes the natural, local shape of the bilayer. While Helfrich~\cite{H} first suggested \eqref{CHexpression} as a model for lipid bilayers, Canham~\cite{C} previously proposed \eqref{CHexpression} with $H_\circ=0$ as a model for red blood cells.  Therefore \eqref{CHexpression} is commonly called the Canham--Helfrich free-energy density.

Most often, $\kappa$ and $\bar\kappa$ are viewed as material parameters, as is $H_\circ$, and this view is adopted here. Whereas $\kappa$ is always positive and can be measured in numerous ways, including, for example, flicker spectroscopy (Brochard and Lennon~\cite{BL}, Schneider, Jenkins and  Webb~\cite{SJW}) and x-ray scattering (Liu and  Nagle~\cite{LN}, Tristram-Nagle and Nagle~\cite{TN}), $\bar\kappa$ is more difficult to quantify. Part of the problem in determining $\bar\kappa$ is related to the Gauss--Bonnet theorem (do Carmo~\cite{doC}), which states that the integral of the Gaussian curvature $K$ over a surface depends only on the topology and boundary of that surface. Granted that $\bar\kappa$ is constant and that the bilayer is closed, the second term on the right-hand side of \eqref{CHexpression} therefore plays a role only in particular processes, such as fusion and fission events, or in multiphase lipid bilayers.

Despite this difficulty, experimental and numerical strategies for obtaining $\bar\kappa$ do exist. As $\kappa$ is relatively easily obtained, it is convenient to specify $\bar\kappa$ through the ratio $\bar\kappa/\kappa$. This ratio is typically found to be negative, with magnitude depending on the constitution of the bilayer. While experiments conducted by Baumgart, Das and Webb~\cite{BDWJ} and by Lorzen, Servuss and Helfrich~\cite{LSH} delivered values of $\bar\kappa/\kappa$ close to $-1$ (namely $-0.9\pm0.38$ and $-0.83\pm0.12$, respectively), experiments conducted by Semrau, Idema, Holtzer, Schmidt and Storm~\cite{SIHSS} delivered values of $\bar\kappa/\kappa$ between $-0.63$ and $-0.31$. Coarse-grained numerical simulations performed by Hu, Brigugli and  Deserno~\cite{HBD} and Hu, de Jong, Marrink and Deserno~\cite{HJMD} yielded values of $-0.95\pm0.1$ and $-1.04\pm0.03$, respectively.  On the basis of a microscopic model for amphiphilic molecules dissolved in water, Chac\'on, Somoza and Tarazona~\cite{CST} obtained a value of $\bar\kappa/\kappa$ of approximately $-1.18$.


A derivation of the Canham--Helfrich free-energy density based on considering the interactions between the lipid molecules that comprise the bilayer was carried out by Seguin and  Fried~\cite{SF}. That derivation provides integral representations for the moduli in terms of a generic interaction potential. In the present work, two potentials are considered, and the integrals are evaluated to obtain $\kappa$ and $\bar\kappa$. Before performing these calculations, two simplifying postulates are imposed: the lipid molecules comprising the bilayer are assumed to be (a) identical and (b) uniformly distributed.  The resulting moduli are described in terms of several parameters.  Besides the molecular number density, these parameters fall into two categories:  those determined by the dimensions of the molecules and those appearing in the interaction potential.  It is shown that the first potential considered, which is an anisotropic Gaussian potential based on those of Berne and Pechukas \cite{BP} and Gay and Berne \cite{GB}, is not able to capture a significant majority of the values for the bending moduli found in the literature. This motivates considering an anisotropic spherocylinder potential introduced recently by Lintuvuori and Wilson \cite{LWilson}, which performs much better.


The paper is organized as follows. Synopses of the salient features of surface geometry and the derivation of Seguin and  Fried~\cite{SF} respectively appear in Sections~\ref{appsurf} and \ref{sectrecap}. Section \ref{sectcalc} contains two subsections, one for each of the potentials considered.  For each potential, the bending moduli are calculated and the resulting expressions are compared to what can be found in the literature.


\section{Geometry of surfaces} \label{appsurf}

Consider a smooth, orientable surface $\cS$ in a three-dimensional Euclidean point space. Let $\bn$ denote a smooth mapping that determines a unit normal at each point of the surface. Given a mapping $h:\cS\longrightarrow\cW$ defined on the surface that takes values in some vector space $\cW$, the surface gradient $\nabla^\cS h$ of $h$ can be defined by
\beqn\label{defsg}
\nabla^\cS_x h:=\nabla_x h^e (\textbf{1}-\bn(x)\otimes\bn(x))\rfa x\in\cS,
\eeqn
where $h^e$ is an extension of $h$ to a neighborhood of $x$ and $\nabla_x h^e$ is the classical three-dimensional gradient of this extension at $x$. Importantly, it can be shown that the definition of the surface gradient is independent of the extension appearing on the right-hand side of \eqref{defsg}.

Of particular interest is the curvature tensor $\bL:=-\nabla^\cS\bn$, the negative of the surface gradient $\nabla^\cS\bn$ of $\bn$, which is a second-order tensor field defined on $\cS$. The tensor $\bL$ is symmetric and has two scalar invariants: the mean curvature $H$ and Gaussian curvature $K$, as defined by
\beqn
\label{H}
H:=\half {\rm tr}\,\bL
\eeqn
and
\beqn
\label{K}
K:=\half[({\rm tr}\,\bL)^2-{\rm tr}(\bL^2)].
\eeqn
If $\lambda_1$ and $\lambda_2$ are the two nontrivial eigenvalues of $\bL$, often called the principle curvatures, then \eqref{H} and \eqref{K} yield
\beqn
H=\half(\lambda_1+\lambda_2)
\qquad\text{and}\qquad
K=\lambda_1\lambda_2.
\eeqn

\section{Recapitulation of the derivation of the Canham--Helfrich free-energy density}\label{sectrecap}


The derivation of Seguin and Fried~\cite{SF} rests on four assumptions:
\begin{enumerate}[I]
\item[(i)] the thickness of the bilayer is small relative to its average diameter;
\item[(ii)] the (phospholipid) molecules can be modeled as one-dimensional rigid rods;
\item[(iii)] the molecules do not tilt relative to the orientation of the bilayer;
\item[(iv)] interactions between the bilayer and the solution are negligible.
\end{enumerate}

Assumption (i), which is often made in models for lipid bilayers (Luisi and  Walde~\cite{LW}), allows the lipid bilayer to be identified with its midsurface $\cS$. This surface may adopt a large variety of shapes; however, being made up of molecules of a finite size, it cannot support arbitrarily large curvatures. Let $\ell$ denote the smallest stable radius of curvature that the bilayer may exhibit. From here on, assume the the bilayer is in a given, fixed configuration $\cS$ at a fixed temperature.

For each leaflet $i=1,2$ of the bilayer, introduce a molecular number density $W_i$ defined on $\cS$ and measured per unit area of $\cS$. Let $\text{d}a_y$ denote the area element on $\cS$. The total number of molecules in leaflet $i=1,2$ is then given by the integral
\beqn
\int_\cS W_i(y)\da.
\eeqn
Taking $W_i$ to be defined on $\cS$ amounts to assuming that the centers of the lipid molecules of both leaflets lie on $\cS$, which is consistent with assuming that the bilayer is thin relative to its average diameter. In general, the number densities of the leaflets may differ.

On the basis of Assumption (ii), the configuration of each molecule in the bilayer may be described by a point on $\cS$ and a unit-vector-valued director, with the point representing the center of the rod and the director representing the orientation of the rod. Without loss of generality, it is assumed that the director tips point toward the headgroups of the molecules. It is further assumed that the interaction between a pair of molecules at two different points on $\cS$ is governed by a potential that depends on a vector connecting the points and the directors at the points and is restricted such that only molecules separated by distances less than some cutoff distance $\ddd$ may interact. Moreover, $\ddd$ is required to be small relative to the smallest radius of curvature $\ell$ the bilayer can support, so that $\ddd\ll\ell$ or, equivalently,
\beqn\label{small}
\epsilon:=\frac{\ddd}{\ell}\ll1.
\eeqn
As will be discussed in the next section, instead of possessing a cutoff distance, some potentials decay rather rapidly as the distance between the interacting molecules increases. For such potentials, it is possible to define an effective cutoff distance beyond which the interaction is negligible and, thus, may be neglected.

Choose points $x$ and $y$ on $\cS$ and consider a molecule at $x$ with orientation $\bd$ and a molecule at $y$ with orientation $\be$. The points $x$ and $y$ should be thought of as coincident with the centers of the molecules. Suppose that interactions between the molecules at $x$ and $y$ are governed by a potential $\Phi$, with dimensions of energy, depending on the vectors $\br=x-y$, $\bd$, and $\be$. Granted that $\Phi$ is frame-indifferent, its dependence on the foregoing quantities must reduce to dependence on the scalars $\br\cdot\br$, $\br\cdot\bd$, $\br\cdot\be$, and $\bd\cdot\be$. Assume that this dependence takes the form
\beqn
\Phi(\br,\bd,\be)=\phi(\epsilon^{-2}\br\cdot\br,\br\cdot\bd,\br\cdot\be,\bd\cdot\be),
\label{Psigeneric}
\eeqn
where $\phi$ satisfies
\beqn\label{ieradius}
\phi(s^2,a,b,c)=0
\quad\text{if $s\ge \ell$ for all $(a,b,c)\in\Real\times\Real\times[-1,1]$.}
\eeqn
The stipulation \eqref{ieradius} ensures that the molecules at $x$ and $y$ interact only if the distance $r=|\br|$ between $x$ and $y$ obeys
\beqn
r<d=\epsilon\ell.
\eeqn
In contrast to $\Phi$, $\phi$ is independent of $d$. Importantly, the interaction energy $\phi$ between two molecules can change on flipping the head group and tails of one of the molecules. Interaction potentials of this form may therefore account for differences between the polarities of the head group and tails of a lipid molecule. Taking $\Phi$ to depend on the cutoff distance $\ddd$ as indicated in \eqref{Psigeneric} is motivated by the work of Keller and  Merchant~\cite{KM}. 

Aside from potentials $\Phi_{11}$ and $\Phi_{22}$ that account for interactions between mol\-e\-cules in each leaflet, it is generally necessary to consider a potential $\Phi_{12}=\Phi_{21}$ that accounts for interactions between molecules belonging to different leaflets. Although the particular forms of the potentials $\Phi_{11}$, $\Phi_{22}$, and $\Phi_{12}=\Phi_{21}$ may differ, they share the same general properties to the extent that they satisfy \eqref{Psigeneric} and \eqref{ieradius}. 

Without loss of generality, orient $\cS$ with a unit-normal field that points into the region adjacent to the head groups of leaflet 1 and denote that field by $\bn$. On the basis of Assumption (iii), it follows that the directors of molecules in leaflets $1$ and $2$ coincide with $\bn$ and $-\bn$, respectively. Bearing in mind the cutoff property \eqref{ieradius}, define $\cS_\ddd(x)$ by
\beqn
\cS_\ddd(x):=\{y\in\cS:\abs{x-y}\leq\ddd\}.
\eeqn
Seguin and  Fried~\cite{SF} argued that the interactions between the lipid molecules making up the bilayer contribute to the free-energy density $\psi$ through the four terms\footnote{Following the lead of Keller and  Merchant~\cite{KM}, Seguin and Fried~\cite{SF} scaled the integrals in \eqref{edtt}--\eqref{edtt4} by $\epsilon^{-2}$. However, upon evaluating these integrals for a particular potential, it transpires that the results scale more appropriately if that scale factor is not introduced.}
\begin{align}
\label{edtt}\psi_{11}(x)&:=\half\int_{\cS_\ddd(x)}\Phi_{11}(x-y,\bn(x),\bn(y))W_1(x)W_1(y)\da,\\
\label{edtt2}\psi_{22}(x)&:=\half\int_{\cS_\ddd(x)}\Phi_{22}(x-y,-\bn(x),-\bn(y))W_2(x)W_2(y)\da,\\
\label{edtt3}\psi_{12}(x)&:=\half\int_{\cS_\ddd(x)}\Phi_{12}(x-y,\bn(x),-\bn(y))W_1(x)W_2(y)\da,\\
\label{edtt4}\psi_{21}(x)&:=\half\int_{\cS_\ddd(x)}\Phi_{21}(x-y,-\bn(x),\bn(y))W_2(x)W_1(y)\da.
\end{align}
The integral in \eqref{edtt} represents the contribution to the free-energy density coming from the interactions between the molecules in leaflet 1 at $x$ and all other molecules in leaflet 1. The integral in \eqref{edtt2} is an analogous contribution involving leaflet 2. The integral in \eqref{edtt3} accounts for the interactions between the molecules in leaflet 1 at $x$ and all other molecules in leaflet 2. The integral in \eqref{edtt4} is analogous to that in \eqref{edtt3}, but with the roles of the two leaflets interchanged.  The factors of one-half in \eqref{edtt}--\eqref{edtt4} ensure that the energy is not double counted. By Assumption (iv), these integrals sum to yield the net free-energy density 
\beqn\label{energydecomp}
\psi=\psi_{11}+\psi_{22}+\psi_{12}+\psi_{21}.
\eeqn

On substituting \eqref{edtt}--\eqref{edtt4} into the right-hand side of \eqref{energydecomp}, $\psi$ can be expanded in powers of $\epsilon$ up to order $\epsilon^4$ with the objective of capturing dependence on the curvature of $\cS$. This expansion takes the form
\beqn\label{mainresult}
\psi=\psi_0+2\kappa(H-H_\circ)^2+\bar\kappa(K-K_\circ),
\eeqn
where $\psi_0$, $\kappa$, $\bar\kappa$, $H_\circ$, and $K_\circ$ are given in terms of $\Phi_{ij}$ and $W_i$.  Here $K_\circ$ is the spontaneous Gaussian curvature.  The homogeneous contribution $\psi_0$ to $\psi$ is of order $\epsilon^2$ and the splay and saddle-splay moduli $\kappa$ and $\bar\kappa$ are of order $\epsilon^4$. Terms of order $\epsilon^4$ are neglected.  A factor of $\epsilon^2$ appears in all terms on the right-hand side of \eqref{mainresult} because the integrals in \eqref{edtt}--\eqref{edtt4} are over a surface with area of order $\epsilon^2$. The moduli $\kappa$ and $\bar\kappa$ contain another factor of $\epsilon^2$ because they stem from the first nontrivial term in the Taylor expansion. A detailed derivation of \eqref{mainresult} is provided by Seguin and Fried \cite{SF}.  The quantities $\psi_0$, $\kappa$, $\bar\kappa$, $H_\circ$, and $K_\circ$ generally depend on the point $x$ in $\cS$ through the molecular number densities $W_i$, $i=1,2$. Thus, $\psi$ may depend on $x$ through not only through the mean and Gaussian curvatures of $\cS$ at $x$ but also through the values of splay and saddle-splay moduli and the spontaneous mean and Gaussian curvatures at $x$. As Seguin and  Fried~\cite{SF} mentioned, the term $\psi_0$ is independent of the shape of the membrane and is commonly neglected in the Canham--Helfrich free-energy density, although Helfrich \cite{H}, for example, did include and discuss it. However, due to implicit dependence of $W_i$ on temperature, concentration, and relevant electromagnetic fields, that term encompasses their effects.  Moreover, $\kappa$ and $\bar\kappa$ depend on these influences implicitly through $W_i$.

Suppose now that:
\begin{enumerate}
\item the molecules of each leaflet are uniformly distributed and the distribution in both leaflets is identical;
\item all of the molecules comprising the bilayer are identical.
\end{enumerate}
As a consequence of Item 1, there is a constant $W$ such that
\beqn
\label{Ione}
W=W_1(x)=W_2(x)\rfa x\in\cS.
\eeqn
Further, as a consequence of Item 2, there is a potential $\Phi$ such that 
\beqn
\label{Itwo}
\Phi=\Phi_{11}=\Phi_{22}=\Phi_{12}=\Phi_{21}.
\eeqn
Granted \eqref{Ione} and \eqref{Itwo}, the spontaneous curvatures vanish and \eqref{mainresult} takes the form
\beqn\label{mainresult2}
\psi=\psi_0+2\kappa H^2+\bar\kappa K.
\eeqn

To provide detailed expressions for $\psi_0$, $\kappa$, and $\bar\kappa$, it is convenient to first introduce the notational conventions
\beqn
\label{appsi0}
\phi_{,0}(s,a):=\phi(s^2,0,0,a)
\eeqn
and
\beqn
\label{appsi}
\phi_{,k}(s,a):=\frac{\partial\phi(\xi_1,\xi_2,\xi_3,\xi_4)}{\partial\xi_k}
\bigg|_{(\xi_1,\xi_2,\xi_3,\xi_4)\mskip1.5mu=\mskip1.5mu(s^2,0,0,a)},
\qquad
k\in\{1,2,3,4\}.
\eeqn
In view of \eqref{Ione} and \eqref{Itwo}, the term $\psi_0$ in \eqref{mainresult2} is given by
\beqn
\label{psi0}
\psi_0:=2\pi\epsilon^2W^2\int_0^\ell\big[\phi_{,0}(r,1)+\phi_{,0}(r,-1)\big]r\dr
\eeqn
and the bending moduli $\kappa$ and $\bar\kappa$ are
\beqn
\label{kappa}\kappa:=B+C
\eeqn
and
\beqn
\label{kappabar}\bar\kappa:=-B,
\eeqn
with $B$ and $C$ defined according to
\beqn
\label{B} 
B:=\pi\epsilon^4W^2\int_0^\ell[\phi_{,0}(r,1)-\phi_{,4}(r,1)+\phi_{,0}(r,-1)+\phi_{,4}(r,-1)]r^3\dr
\eeqn
and
\beqn
\label{C}
C:=\frac{3\pi\epsilon^4}{8}W^2\int_0^\ell[\phi_{,1}(r,1)+\phi_{,1}(r,-1)]r^5\dr.
\eeqn

%
%

The signs of $\kappa$ and $\bar\kappa$ are sometimes set by the signs of $\phi_{,0}$, $\phi_{,1}$, and $\phi_{,4}$, which are determined by the properties of the potential $\phi$. In particular, the sign of $\phi_{,1}$ is linked to whether $\phi$ is attractive or repulsive:
\begin{itemize}
\item if the potential is attractive, then 
\beqn
\phi_{,1}(r,\pm 1)\geq0 \quad\text{for all $r$};
\eeqn
\item if the potential is repulsive, then 
\beqn
\phi_{,1}(r,\pm 1)\leq0\quad\text{for all $r$}.
\eeqn
\end{itemize}
Potentials may, of course, possess attractive and repulsive domains, as is the case for the Gay--Berne \cite{GB} potential $\phi^{\rm GB}$, for which the sign of $\phi^{\rm GB}_{,1}(r,\pm 1)$ depends on $r$. If $\phi$ obeys \eqref{ieradius} and is attractive (repulsive), then $\phi_{,0}\leq0$ ($\phi_{,0}\geq0$).  

Evaluating the potential and its partial derivatives at the values $(s^2,0,0,\pm1)$ (see \eqref{appsi0}--\eqref{psi0} and \eqref{B}--\eqref{C}) is akin to considering side-by-side configurations for the molecules. In particular, the sign of $\phi_{,4}(r,\pm1)$ is linked to whether such a configuration is favorable:
\begin{itemize}
\item if side-by-side configurations are favorable, then 
\beqn\label{sbsfav}
\phi_{,4}(r,1)\leq0\quad\text{and}\quad\phi_{,4}(r,-1)\geq0\quad\text{for all $r$};
\eeqn
\item if side-by-side configurations are unfavorable, then 
\beqn
\phi_{,4}(r,1)\geq0\quad\text{and}\quad\phi_{,4}(r,-1)\leq0\quad\text{for all $r$}.
\eeqn
\end{itemize}
In view of \eqref{kappa}--\eqref{C} and the foregoing observations, $\kappa\geq0$ and $\bar\kappa\leq0$ for a repulsive potential that favors side-by-side configurations but $\kappa\leq0$ and $\bar\kappa\geq0$ for an attractive potential that does not favor side-by-side configurations. Since $\kappa\leq0$ is physically unsound, it is unreasonable to use an attractive potential that does not favor side-by-side configurations. If the potential is neither attractive nor repulsive, then determining the signs of $\kappa$ and $\bar\kappa$ is more involved.


\section{Calculations using particular potentials}\label{sectcalc}

In this section, the bending moduli $\kappa$ and $\bar\kappa$ are computed using two potentials, and the results are discussed. Prior to this, a few words on the choice of the potentials seem appropriate.

The literature is replete with potentials designed to describe the interactions between molecules. Of interest here are potentials appropriate to molecules resembling one-dimensional rods possessing an axis of symmetry and being relatively long in that direction.

It is possible to consider two categories of pair potentials: those with hard cores and those with soft cores. The energy of a hard-core potential becomes infinite as the distance between the interacting molecules approaches zero. This property reflects the impossibility of molecular overlap. For a soft-core potential, the energy tends to a finite value as the distance between the molecules approaches zero.

In the present work, only soft-core potentials are considered. This is because the model for the lipid bilayer considered here is continuous rather than discrete. To compute the moduli $\kappa$ and $\bar\kappa$, interactions between molecules arbitrarily close together must be considered.  Since hard-core potentials blow up as molecules become arbitrarily close, using a hard-core potential would result in an infinite bending moduli, which is certainly not useful.

\subsection{An anisotropic Gaussian potential}\label{sectchoice}

%

The first potential considered will exhibit a multiplicative decomposition in which one factor, referred to as the strength parameter, is independent of the distance between the molecules, while the other factor depends on the distance and tends to zero as it approaches infinity.  To illustrate the properties of such a potential, consider axisymmetric particles at $x$ and $y$ with respective directors $\bd$ and $\be$. Introduce the unit vector
\beqn\label{rnotation}
\hat\br=\frac{\br}{r},
\qquad
r=\abs{\br},
\eeqn
in the direction of $\br=x-y\ne\bf0$. A potential $\Phi$ manifesting the aforementioned multiplicative decomposition can be written in the form
\beqn\label{phidecomp}
\Phi(\br,\bd,\be)=S(\hr,\bd,\be)\Sigma(\br,\bd,\be),
\eeqn
where $S$ is the strength parameter\footnote{The strength parameter is commonly denoted by $\epsilon$, but that symbol is already used here in different context.}, and $\Sigma$ satisfies
\beqn\label{decay}
\lim_{r\rightarrow\infty}\Sigma(\br,\bd,\be)=0.
\eeqn
%

For particles with an axis of symmetry, it is common to consider a ``Gaussian'' potential. A short survey of such potentials is given by Walmsley~\cite{W}, who observes that for a Gaussian potential it is common to choose $\Sigma$ to have the form
\beqn\label{Sigmaform}
\Sigma(\br,\bd,\be)=f(r^{-1}\sigma(\hr,\bd,\be)),
\eeqn
where $\sigma$ is the range parameter. Not all potentials are of the type described by \eqref{Sigmaform}. A noteworthy exception is due to Gay and  Berne~\cite{GB}, who use $\sigma$ but choose an expression for $\Sigma$ not contained within the class considered by Walmsley~\cite{W}.

For the strength parameter $S$, an expression proposed by Gay and  Berne~\cite{GB} is used. This expression has the form
\beqn\label{epsilon}
S(\hr,\bd,\be):=S_0 S_1(\bd,\be)^\nu S_2(\hr,\bd,\be)^\mu,
\eeqn
where $S_0$, $\nu$, and $\mu$ are parameters to be chosen and $S_1$ and $S_2$ are given by
\beqn
S_1(\bd,\be):=\frac{1}{\sqrt{1-\chi^2(\bd\cdot\be)^2}}
\eeqn
and
\beqn
S_2(\br,\bd,\be):=1-\frac{\chi'}{2}\left( \frac{(\hr\cdot\bd+\hr\cdot\be)^2}{1+\chi'(\bd\cdot\be)}+\frac{(\hr\cdot\bd-\hr\cdot\be)^2}{1-\chi'(\bd\cdot\be)}\right).
\eeqn
Here, $\chi$ and $\chi'$ are defined in accord with
\beqn
\label{chichiprime}
\chi:=\frac{\rho^2-1}{\rho^2+1}
\qquad\text{and}\qquad
\chi':=\frac{1-(\epsilon_E/\epsilon_S)^{1/\mu}}{1+(\epsilon_E/\epsilon_S)^{1/\mu}},
\eeqn
where $\rho$ is the aspect ratio (length divided by {diameter) of a molecule and $\epsilon_E$ and $\epsilon_S$ are the strength parameters for end-to-end and side-to-side interactions. For slender molecules, $\rho$ is closer to infinity than to $0$ and, thus, $\chi$ is closer to $1$ than $0$.

Following Berne and  Pechuckas~\cite{BP}, $\Sigma$ is taken to be of the form
\beqn\label{Sigma}
\Sigma(\br,\bd,\be):=
\exp\left({-\frac{r^2}{\sigma(\hr,\bd,\be)^2}}\right),
\eeqn
with the range parameter $\sigma$ given by
\beqn
\sigma(\hr,\bd,\be):=\frac{\sigma_0}{\displaystyle\left[1-\frac{\chi}{2}\left( \frac{(\hr\cdot\bd+\hr\cdot\be)^2}{1+\chi\,\bd\cdot\be}+\frac{(\hr\cdot\bd-\hr\cdot\be)^2}{1-\chi\,\bd\cdot\be}\right)\right]^{1/2}},
\eeqn
where, on using $D$ to denote the diameter of a molecule,
\beqn
\sigma_0:=\sqrt{2}D.
\eeqn
The particular choice \eqref{Sigma} of $\Sigma$ leads to a potential that does not have a cutoff distance. However, as Earl~\cite{Earl} observes, since the potential decays exponentially as the ratio $r/\sigma_0$ becomes large, it is reasonable to define an effective cutoff distance
\beqn\label{effcutdist}
d=3 \sigma_0.
\eeqn

Looking at \eqref{psi0}--\eqref{C}, the interaction potential $\phi$ only enters $\psi_0$ and the moduli $\kappa$ and $\bar\kappa$ through the expressions
\begin{align}
\label{phi0}\phi_{,0}(r,\pm 1)&=\frac{S_0}{(1-\chi^2)^{\nu/2}}
\exp\left({-\frac{d^2r^2}{\ell^2\sigma_0^2}}\right),
\\
\label{phi1}\phi_{,1}(r,\pm 1)&=-\frac{S_0d^2}{(1-\chi^2)^{\nu/2}\ell^2\sigma_0^2}
\exp\left({-\frac{d^2r^2}{\ell^2\sigma_0^2}}\right),
\\
\label{phi4}\phi_{,4}(r,\pm 1)&=\pm\frac{S_0\chi^2 \nu}{(1-\chi^2)^{1+\nu/2}}
\exp\left({-\frac{d^2r^2}{\ell^2\sigma_0^2}}\right).
\end{align}
From \eqref{phi0}--\eqref{phi4} it is clear that $\psi_0$, $\kappa$, and $\bar\kappa$ can be determined without reference to the values of $\mu$ and $\chi'$ but are, however, influenced by $\nu$, $S_0$, $\sigma_0$, and $\chi$. As it transpires that working with $\sigma_0$ and $\chi$ is more convenient than $\rho$ and $D$, only $\sigma_0$ and $\chi$ will be used hereafter.

Notice that of $\phi_{,0}(r,a)$, $\phi_{,1}(r,a)$, and $\phi_{,4}(r,a)$, only $\phi_{,4}(r,a)$ is sensitive to whether $a=1$ or $a=-1$. This observation can be interpreted once it is taken into consideration that $\phi_{,2}(r,a)=\phi_{,3}(r,a)=0$ and, as Seguin and  Fried \cite{SFLC} found, that the force $\bff$ and couple $\bC$ exerted on a molecule at $x$ with director $\be$ from a molecule at $x+\br$ with director $\bd$ are given respectively by
\beqn
\bff=-2\epsilon^{-2}\phi_{,1}\br-\phi_{,2}\bd-\phi_{,3}\be
\eeqn
and
\beqn
\bC=-\phi_{,2}\bd\wedge\br-\phi_{,4}\bd\wedge\be.
\eeqn
The insensitivity of $\phi_{,0}$ and $\phi_{,1}$ to whether $a=1$ or $a=-1$ implies that the interaction energy and forces between molecules in side-by-side configurations are the same regardless of how the molecules are oriented relative to each other. However, since $\phi_{,4}(r,a)$ is sensitive to whether $a=1$ or $a=-1$, in side-by-side configurations the couple exerted by one molecule on another is influenced by molecular orientation.

The sign of the parameter $\nu$ deserves some discussion. The absolute value of $\nu$ determines the extent to which the strength of the interaction between molecules is affected by their relative orientation. Whereas parallel configurations of molecules are preferred for $\nu<0$, perpendicular configurations are preferred for $\nu>0$. Molecular orientation does not influence the strength of the interaction if the orientation strength vanishes. Since the lipid molecules comprising the bilayer prefer to be parallel, a negative value of the orientation strength is appropriate.

Notice that the potential specified in this subsection is repulsive.  Using this potential begs the question as to what causes the lipid molecules to self-assemble into a bilayer. That process emanates from interactions between the molecules and the host solution, interactions that cause the molecules to arrange themselves to shield their tails from the solution. The model utilized here does not address this phenomenon. Rather, it presumes that the molecules are already configured in the shape of a bilayer and that the energy of this structure is due to the interaction between the molecules comprising the bilayer.



Using \eqref{phi0}--\eqref{phi4} in the integrals appearing on the right-hand sides of the expression \eqref{psi0} for the zero curvature contribution $\psi_0$ to the free-energy density and the definitions \eqref{B} and \eqref{C} of the quantities $B$ and $C$ needed to determine the bending moduli $\kappa$ and $\bar\kappa$ through \eqref{kappa} and \eqref{kappabar} leads to integrals that can be evaluated in closed form using identities provided by Albano, Amdeberhan, Beyerstedt and Moll \cite{AABM}. Specifically, with \eqref{phi0}, using \eqref{small}, the right-hand side of \eqref{psi0} gives
\beqn
2\pi\epsilon^2W^2\int_0^\ell\big[\phi_{,0}(r,1)+\phi_{,0}(r,-1)\big]r\dr
=\frac{2\pi S_0\sigma_0^2W^2}{(1-\chi^2)^{\nu/2}}
\bigg[1-\exp\bigg(-\frac{d^2}{\sigma_0^2}\bigg)\bigg],
\label{term1}
\eeqn
while, with \eqref{phi0} and \eqref{phi4}, the right-hand side of \eqref{B} gives
\begin{multline}
\pi\epsilon^4W^2\int_0^\ell[\phi_{,0}(r,1)-\phi_{,4}(r,1)+\phi_{,0}(r,-1)+\phi_{,4}(r,-1)]r^3\dr
\\[4pt]
=\frac{\pi(1-\chi^2-\chi^2\nu)S_0\sigma_0^4W^2}{(1-\chi^2)^{1+\nu/2}}
\bigg[1-\bigg(1+\frac{d^2}{\sigma_0^2}\bigg)\exp\bigg(-\frac{d^2}{\sigma_0^2}\bigg)\bigg],
\label{term2}
\end{multline}
and, with \eqref{phi1}, the right-hand side of \eqref{C} gives
\begin{multline}
\frac{3\pi\epsilon^4W^2}{8}\int_0^\ell[\phi_{,1}(r,1)+\phi_{,1}(r,-1)]r^5\dr
\\[4pt]
=-\frac{3\pi S_0\sigma_0^4W^2}{4(1-\chi^2)^{\nu/2}}
\bigg[1-\bigg(1+\frac{d^2}{\sigma_0^2}
+\frac{d^4}{2\sigma_0^4}\bigg)\exp\bigg(-\frac{d^2}{\sigma_0^2}\bigg)\bigg].
\label{term3}
\end{multline}
On invoking the definition \eqref{effcutdist} of the effective cutoff distance, the terms in square brackets on right-hand sides of \eqref{term1}--\eqref{term3} are all well approximated by 1. With this in mind, \eqref{psi0} and \eqref{term1} yield
\beqn
\psi_0=\frac{2\pi S_0\sigma^2_0\ell^2W^2}{(1-\chi^2)^{\nu/2}},
\eeqn
\eqref{kappa}, \eqref{B}, \eqref{C}, \eqref{term2}, and \eqref{term3} yield
\beqn
\kappa=\frac{\pi(1-\chi^2-4\chi^2\nu)S_0\sigma^4_0\ell^4W^2}{4(1-\chi^2)^{1+\nu/2}},
\label{kappaev}
\eeqn
and, finally, \eqref{kappabar}, \eqref{C}, and \eqref{term3}, $\bar\kappa$ yield
\beqn
\bar\kappa=-\frac{\pi(1-\chi^2-\chi^2\nu)S_0\sigma^4_0\ell^4W^2}{(1-\chi^2)^{1+\nu/2}}.
\label{barkappaev}
\eeqn
%
%
%
Since $0<\chi<1$ and $\nu<0$, \eqref{kappaev} and \eqref{barkappaev} imply that $\kappa>0$ and $\bar\kappa<0$.  In view of the discussion in the final paragraph of Section~\ref{sectrecap}, this is unsurprising as the chosen potential is repulsive and favors side-by-side configurations. The ratio $\bar\kappa/\kappa$ of the bending moduli is given by
\beqn\label{ratioev}
\frac{\bar\kappa}{\kappa}=-\frac{4(1-\chi^2-\chi^2\nu)}{1-\chi^2-4\chi^2\nu}.
\eeqn

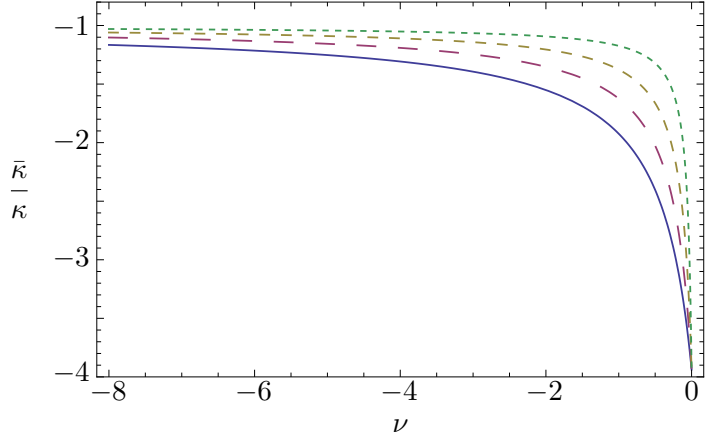
\begin{figure*}[t]
\begin{center}
{\includegraphics[width=3.2in]{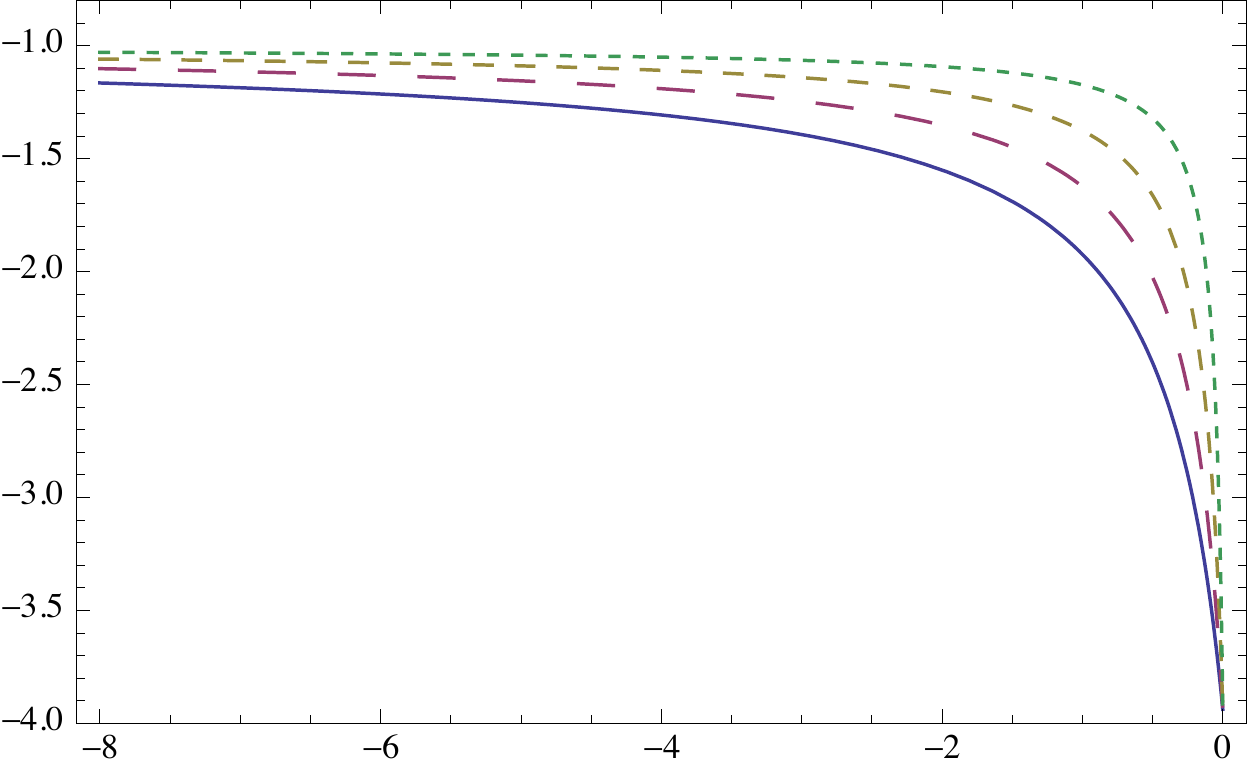}}
\put(-264,68.6){$\displaystyle\frac{\bar\kappa}{\kappa}$}
\put(-118.5,-20){$\nu$}
\put(-247,130.5){$-1$}
\put(-247,87){$-2$}
\put(-247,43.0){$-3$}
\put(-247,-0.5){$-4$}
\put(-233,-8){$-8$}
\put(-178,-8){$-6$}
\put(-123,-8){$-4$}
\put(-68,-8){$-2$}
\put(-8,-8){$0$}
\vspace{-8pt}
\end{center}
\caption{Plot of the ratio $\bar\kappa/\kappa$ as a function of the orientation strength $\nu$ for different choices of the parameter $\chi$ defined in terms of the molecular aspect ratio $\rho$ in \eqref{chichiprime}$_1$. The solid, long-dashed, medium-dashed, and short-dashed lines correspond respectively to $\chi=0.6$, $\chi=0.7$, $\chi=0.8$, and $\chi=0.9$ or, equivalently, $\rho=2.0$, $\rho=2.4$, $\rho=3.0$, and $\rho=4.4$.}
\end{figure*}

Plots of the ratio $\bar\kappa/\kappa$ as a function of $\nu<0$ are provided in Figure 1 for various values of $\chi$. As a consequence of \eqref{chichiprime}$_1$, there is a one-to-one correspondence between the value of $\chi$ and the aspect ratio $\rho$ of the lipid molecules, which increases as $\chi$ approaches $1$. For $\chi<1$, \eqref{ratioev} implies that $\bar\kappa/\kappa$ may take any value in the interval $(-4,-1)$. 

The result \eqref{kappaev} agrees with what appears in the literature in the sense that the parameter $S_0$ may be selected to ensure that the resulting value for the modulus $\kappa$ is consistent with those obtained through experimental measurements and numerical simulations.  This is because $S_0$ can be chosen so that $\kappa$ takes any desired positive value.  However, the result \eqref{ratioev} does not agree with many of the available values. Besides covering half of the values obtained by Hu, de Jong, Marrink and Deserno~\cite{HJMD} and the value found by Chac\'on, Somoza and Tarazona~\cite{CST}, \eqref{ratioev} cannot be made to match the values of $\bar\kappa/\kappa$ found in the other works mentioned in the introduction.  This motivates considering a different potential, which is done next.


\subsection{Lintuvuori--Wilson potential}

Lintuvuori and Wilson \cite{LWilson} recently introduced an anisotropic soft-core potential for pairwise spherocylinder-spherocylinder interactions. In addition to incorporating repulsive and attractive domains, this potential possesses a definite cutoff distance.

The strength $S_\text{a}$ of the attractive branch of the Lintuvuori--Wilson potential, which depends on $\hat\br$ and the orientations $\bd$ and $\be$ as well, is given by
\beqn\label{Sa}
S_\text{a}(\hat\br,\bd,\be)
=S_1-\eta_1P_2(\bd\cdot\be)-\eta_2[P_2(\hat\br\cdot\bd)+P_2(\hat\br\cdot\be)],
\eeqn
where $P_2(x)=\frac{1}{2}(3x^2-1)$ is the second Legendre polynomial. Whereas $S_1$ controls the strength of the attractive part of the interaction independent of the orientations of the molecules, $\eta_1$ and $\eta_2$ dictate the extent to which the orientations influence the magnitude of the attractive interaction.

Given $\hat\br$, $\bd$, and $\be$, an interaction potential between rod-like molecules can be considered as a function solely of $r$.  If the potential has a cutoff distance, then there is a  $d_\text{o}$, possibly dependent on $\hat\br$, $\bd$, and $\be$, such that the potential vanishes for $r$ larger than $d_\text{o}$. For reasons to be explained below, Lintuvuori and Wilson \cite{LWilson} choose $d_\text{o}$ to be of the form
\beqn\label{dcut}
d_\text{o}(\hat\br,\bd,\be)=1
+\frac{1}{\sqrt{2S_\text{a}(\hat\br,\bd,\be)}}.
\eeqn
The cutoff distance $d$ introduced in the paragraph containing \eqref{small} is the maximum of $d_\text{o}$ over all possible unit vectors $\hat\br$, $\bd$, and $\be$.  It is possible to think of $d_\text{o}$ as an orientation-dependent cutoff distance and $d$ as a global cutoff distance.

According to Lintuvuori and Wilson \cite{LWilson}, the interaction energy between two spherocylindrical molecules of length $L$ is given by
\beqn
\Phi(\br,\bd,\be):=\left\{
\begin{array}{cc}
S_0(1-\frac{r}{L})^2+S_0\xi(\hat\br,\bd,\be), &\hspace{.1in} \frac{r}{L}<1,\\[6pt]
\begin{array}{c}
S_0(1-\frac{r}{L})^2-S_0S_\text{a}(\hat\br,\bd,\be)(1-\frac{r}{L})^4\\[2pt]
+S_0\xi(\hat\br,\bd,\be),
\end{array}
&\hspace{.1in} 1\leq \frac{r}{L}< d_\text{o}(\hat\br,\bd,\be),\\[10pt]
0, &\hspace{.1in}  d_\text{o}(\hat\br,\bd,\be)\leq \frac{r}{L},
\end{array}
\right.
\eeqn
where $\xi$ has the form
\beqn
\xi(\hat\br,\bd,\be)=-\frac{1}{4S_\text{a}(\hat\br,\bd,\be)}
\eeqn
and is chosen to ensure that $\Phi$ is continuously differentiable and the same rationale underlies the chosen form \eqref{dcut} of $d_\text{o}$}.  The parameter $S_0$ controls the overall strength of the interaction.


As was noted at the end of Section 3, to calculate $\psi_0$, $\kappa$, and $\bar\kappa$ using \eqref{psi0}, \eqref{kappa}, and \eqref{kappabar} it is sufficient to restrict attention to side-by-side configurations of the molecules.  For such configurations, $\hat\br\cdot\bd=\hat\br\cdot\be=0$, in which case it can be shown that \eqref{Sa} can be replaced by
\beqn
S_\text{a}(\hat\br,\bd,\be)=S_1-\eta_1P_2(\bd\cdot\be)-\eta_2
\eeqn
without effecting the values of $\kappa$ and $\bar\kappa$. It thus suffices to consider the single parameter $S_2:=S_1-\eta_2$ rather than the two parameters $S_1$ and $\eta_2$. From \eqref{dcut}, it is evident that the potential is well-defined only when $S_\text{a}$ is positive and, therefore, when
\beqn
\label{S2etaineqalities}
S_2> 0\qquad \text{and}\qquad -2S_2< \eta_1< S_2.
\eeqn
Moreover, it seems reasonable to only consider a potential that makes side-by-side configurations favorable---that is, to assume that \eqref{sbsfav} holds.  The Lintuvuori--Wilson potential favors side-by-side configurations exactly when $\eta_1\geq0$.

\begin{figure*}[t]
\begin{center}
{\includegraphics[width=3.2in]{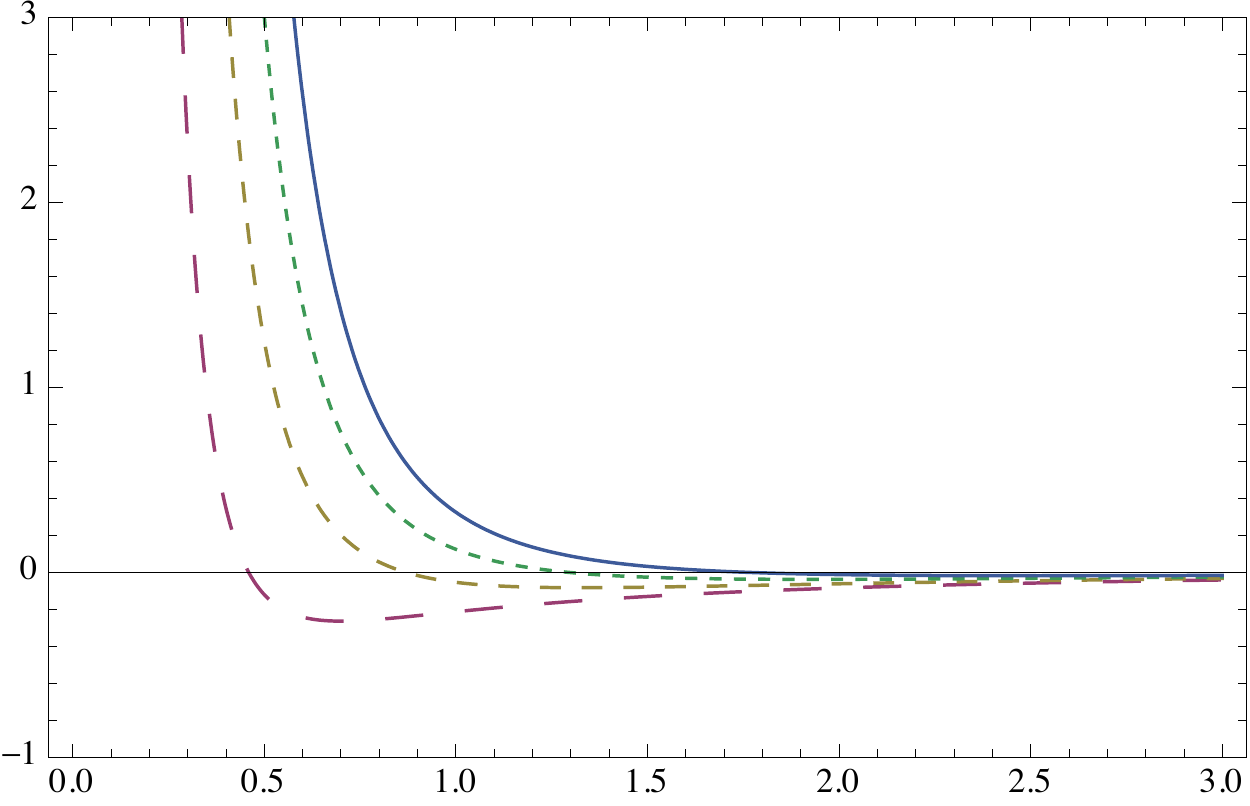}}
\put(-290,75){$\displaystyle\frac{\kappa}{S_0L^4W^2}$}
\put(-118.5,-23){$S_2$}
\put(-239,139){$3$}
\put(-239,104){$2$}
\put(-239,69){$1$}
\put(-239,34){$0$}
\put(-245,-0.5){$-1$}
\put(-228,-8){$0$}
\put(-196,-8){$0.5$}
\put(-159,-8){$1.0$}
\put(-122,-8){$1.5$}
\put(-85.5,-8){$2.0$}
\put(-49,-8){$2.5$}
\put(-12.5,-8){$3.0$}
\vspace{-8pt}
\end{center}
\caption{Plot of $\frac{\kappa}{S_0L^4W^2}$ as a function of $S_2$ for different choices of the parameter $\eta_1$. The solid, long-dashed, medium-dashed, and short-dashed lines correspond respectively to $\eta_1=0.8$, $\eta_1=0.6$, $\eta_1=0.4$, and $\eta_1=0.2$.}
\end{figure*}
\begin{figure*}[t]
\begin{center}
{\includegraphics[width=3.2in]{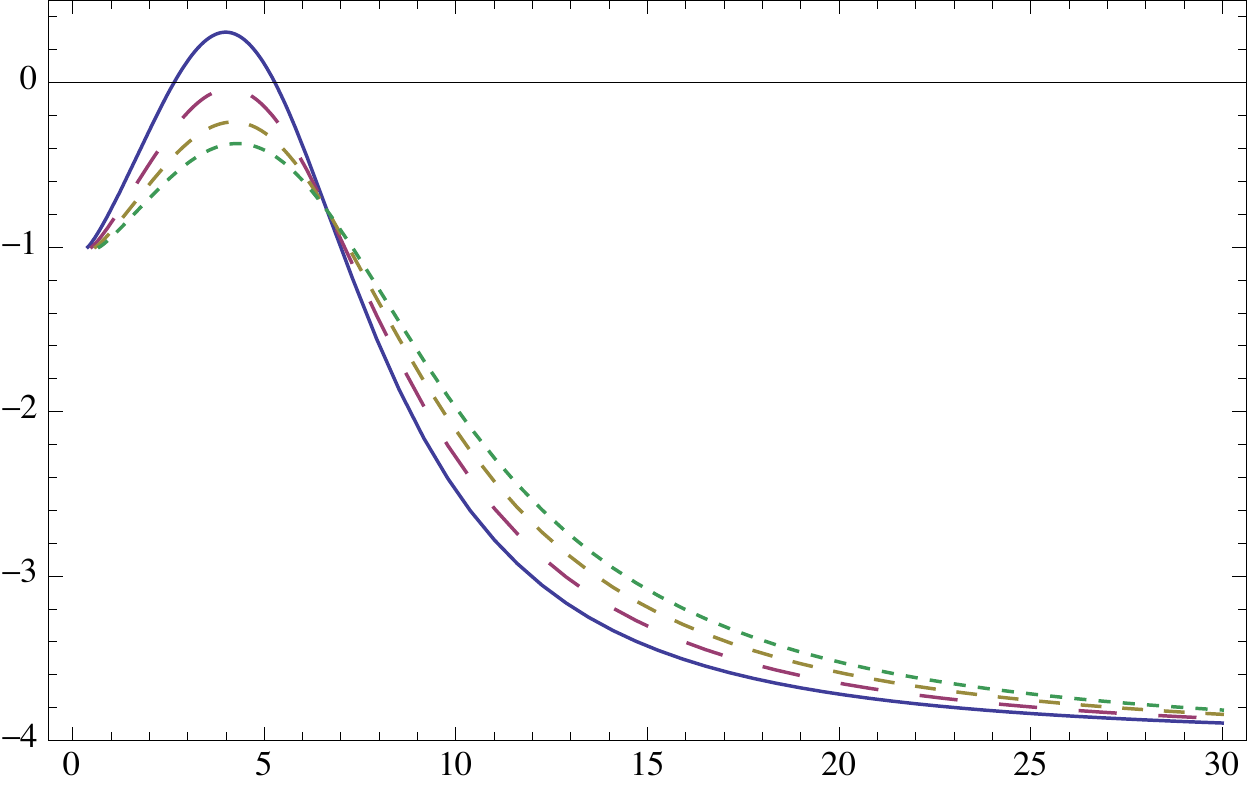}}
\put(-270,68.6){$\displaystyle\frac{\bar\kappa}{\kappa}$}
\put(-118.5,-28){$S_2$}
\put(-240,123){$0$}
\put(-247,92){$-1$}
\put(-247,60){$-2$}
\put(-247,29){$-3$}
\put(-247,-2){$-4$}
\put(-227,-8){$0$}
\put(-190.5,-10){$5$}
\put(-156,-10){$10$}
\put(-120,-10){$15$}
\put(-83,-10){$20$}
\put(-46,-10){$25$}
\put(-9,-10){$30$}
\vspace{-8pt}
\end{center}
\caption{Plot of the ratio $\bar\kappa/\kappa$ as a function of $S_2$ for different choices of the parameter $\eta_1$. The solid, long-dashed, medium-dashed, and short-dashed lines correspond respectively to $\eta_1=0.4$, $\eta_1=0.5$, $\eta_1=0.6$, and $\eta_1=0.7$.}
\end{figure*}
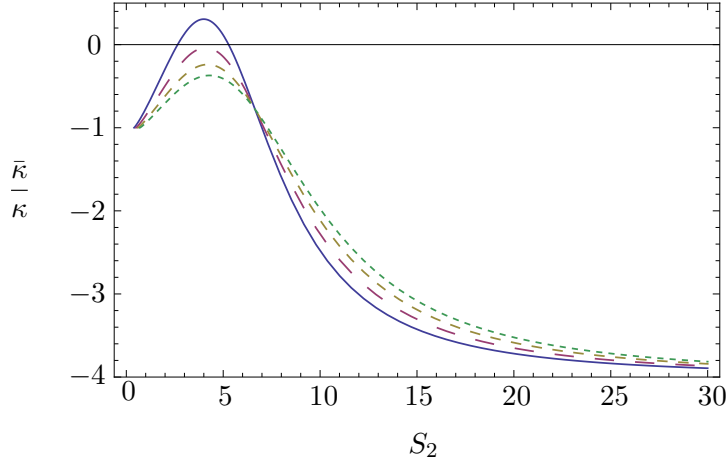

In view of \eqref{dcut}--\eqref{S2etaineqalities}, and making use of Mathematica \cite{Wolf}, using the Lintuvuori--Wilson potential to determine \eqref{psi0} results in 
\beqn
\psi_0= \bigg(20-\frac{5}{(S_2-\eta_1)^2}-\frac{16\sqrt{2}}{(S_2-\eta_1)^{3/2}}-\frac{30}{S_2-\eta_1}\bigg )\frac{\pi S_0L^2W^2}{60(S_2-\eta_1)}
\eeqn
and the splay modulus $\kappa$ determined from \eqref{kappa} results in
\beqn
\kappa=f(S_2,\eta_1)S_0L^4W^2,
\eeqn
where $f$ is a function of $S_2$ and $\eta_1$ too lengthy to warrant inclusion here. For values of $\eta_1$ greater than approximately $0.13$, the function $f(\cdot,\eta_1)$ is positive.  Moreover,
\beqn
\lim_{S_2\rightarrow \eta_1}f(S_2,\eta_1)=\infty\qquad\text{and}\qquad\lim_{S_2\rightarrow \infty}f(S_2,\eta_1)=0.0256
\eeqn
for all $\eta_1>0$. These features of $\kappa$ are made apparent in Figure 2.

The ratio $\bar\kappa/\kappa$ of the moduli determined from \eqref{kappabar} using the Lintuvuori--Wilson potential depends on $S_2$ and $\eta_1$ in a manner depicted in Figure 2. When $S_2$ is close to $\eta_1$, the ratio $\bar\kappa/\kappa$ is close to $1$.  For $S_2$ increasing and each value of $\eta_1$ satisfying $\eta_1\gtrsim 0.13$, the ratio $\bar\kappa/\kappa$ increases to a maximum and then decreases until it levels off at a value of $-4$. The maximum value of the ratio $\bar\kappa/\kappa$ depends on the value of $\eta_2$ and increases as $\eta_2$ decreases. For $\eta_1\lesssim0.13$, $\bar\kappa/\kappa$ exhibits a vertical asymptote in its dependence on $S_2$.  

The physical significance of the requirement $\eta_1\gtrsim 0.13$ needed to ensure that $\kappa$ is positive and that $\bar\kappa/\kappa$ does not blow up remains uncertain.   

As is evident from Figure 3, suitable choices of the parameters $S_2$ and $\eta_1$ allow $\bar\kappa/\kappa$ to take any value in the interval $(-4,0]$.  Thus, suitable choices for these parameters allows for the model put forward by Seguin and Fried \cite{SF} to accurately predict the entire range of experimentally and numerically determined values of $\bar\kappa/\kappa$ found in the literature.

\subsection*{Acknowledgment}

We express thanks to Prof.\ L.\ Mahadevan for recommending that we consider calculating the splay and saddle splay moduli using the Gay--Berne potential and to Dr.~Mohsen Maleki, Dr.~Russell E.\ Todres, and Prof.~Markus Deserno for constructive comments and suggestions.

\end{document}